\documentclass[preprint,12pt, 3p, onecolumn]{elsarticle}

\usepackage{amssymb}
\usepackage{color}

\journal{Journal of Crystal Growth}

\begin{document}
\biboptions{sort&compress}
\begin{frontmatter}
\title{Crystal growth by Bridgman and Czochralski method of the ferromagnetic quantum critical material YbNi$_4$P$_2$}
\author{K.~Kliemt\corref{cor1}} 
 \ead{kliemt@physik.uni-frankfurt.de}
\author{C.~Krellner\corref{cor2}} 
 \address{Kristall- und Materiallabor, Physikalisches Institut, 
 Goethe-Universit\"at Frankfurt, Max-von-Laue Stasse 1, 
 60438 Frankfurt am Main, Germany}
\cortext[cor1]{Corresponding author}

\begin{abstract}
The tetragonal YbNi$_4$P$_2$ is one of the rare examples of compounds 
that allow the investigation of a ferromagnetic quantum critical point. We report in detail on
two different methods which have been used to grow YbNi$_4$P$_2$ single crystals from a self-flux. 
The first, a modified Bridgman method, using a closed crucible system yields needle-shaped single crystals oriented along the $[001]$-direction. 
The second method, the Czochralski growth from a levitating melt, yields large single crystals which can 
be cut in any desired orientation. 
With this crucible-free method, samples without flux inclusions and a resistivity ratio at 1.8 K of RR$_{1.8\rm K}= 17$ have been grown.
\end{abstract}

\end{frontmatter}

\section{Introduction}
\label{}
In the last decades, compounds containing lanthanides (Ln) have been studied due to their large variety 
of interesting physical properties like quantum criticality,
intermediate valence states,  
complex or anisotropic magnetism,  
heavy fermion behaviour 
as well as 
the occurence of unconventional superconductivity 
\cite{Brandt1984, Stewart2001, Stewart2006, Pfleiderer2009}. 
Growing crystals of Yb compounds containing transition metals means dealing with the high vapour pressure of the first 
and the high melting temperature of the latter.
For the crystal growth therefore often a flux method at high temperatures is applied. 
By using a flux it is possible to solve the high-melting elements as well as the elements with low boiling points  
and obtain a melt with a moderate vapour pressure which is suitable for the growth.  
A good overview about the use of metallic fluxes is given in \cite{Canfield1992,Canfield2001}.
In the past, the flux method has been successfully applied for the growth of Ln compounds using indium,
tin or lithium as flux \cite{Onuki2007,Seiro2014,Kliemt2015,Krellner2008,Kanatzidis2005,Jesche2014}.
In some cases, the use of a solvent leads to the formation of unwanted phases, in this regard
the use of a self-flux can be more successful. 
Even when using a flux, the growth temperature often exceeds 1200$^\circ$C. 
Due to the highly volatile and reactive constituents the growth usually is performed in a closed Nb or Ta crucible. Another attempt is the application of inert gas pressure during the growth.
Due to its high reactivity, reports on the growth of phosphorous containing bulk single crystals are rare.
Binary phosphides have been grown by a liquid-encapsulated Czochralski method (InP \cite{Bachmann1975},\cite{Sun2001}),
by chemical vapour phase transport (CuP$_2$ \cite{Kloc1990}) or
under high pressure (CoP$_3$ \cite{Lee2004}).
Ternary phosphides have been grown in a closed crucible from tin flux (LnRu$_2$P$_2$\cite{Fujiwara2011,Kanatzidis2005}) or 
by applying the ACRT Bridgman method (ZnGeP$_2$  \cite{Zhang2011}).

Within this manuscript, we report in detail on the growth of the intermetallic compound YbNi$_4$P$_2$.
Quantum phase transitions that occur at zero temperature are of current interest in solid state 
physics. YbNi$_4$P$_2$ is one of the rare examples of compounds 
that allow the investigation of a ferromagnetic quantum critical point (FM QCP).
Low-temperature measurements of Steppke et al. \cite{Steppke2013} indicate the existence of a FM QCP in 
YbNi$_4$(P$_{1-\rm x}$As$_{\rm x}$)$_2$. For further investigation of this intriguing phenomenon, high quality as well as large 
single crystals are essential.  
YbNi$_4$P$_2$ crystallizes in the  
tetragonal ZrFe$_4$Si$_2$ structure type (P$4_2$/mnm). 
In this rather unexplored structure type, the Yb atoms are located in channels of Ni tetrahedral chains leading to quasi-1D character also of the electronic structure of this compound. 
YbNi$_4$P$_2$ is a heavy fermion compound (T$_K \approx$ 8 K) and orders ferromagnetically below T$_C \approx $150 mK \cite{Krellner2011}.
The magnetic properties were investigated by magnetization measurements \cite{Chikhrij1991, Deputier1997, Krellner2011, Krellner2012}, NMR \cite{Sarkar2012, Sarkar2013} and $\mu$SR \cite{Spehling2012}.
Inelastic neutron scattering on powder was performed to investigate the crystalline electric-field splitting and ferromagnetic fluctuations \cite{Huesges2013, Huesges2015}. 

\begin{figure}[ht]
\centering
\includegraphics[width=0.5\columnwidth]{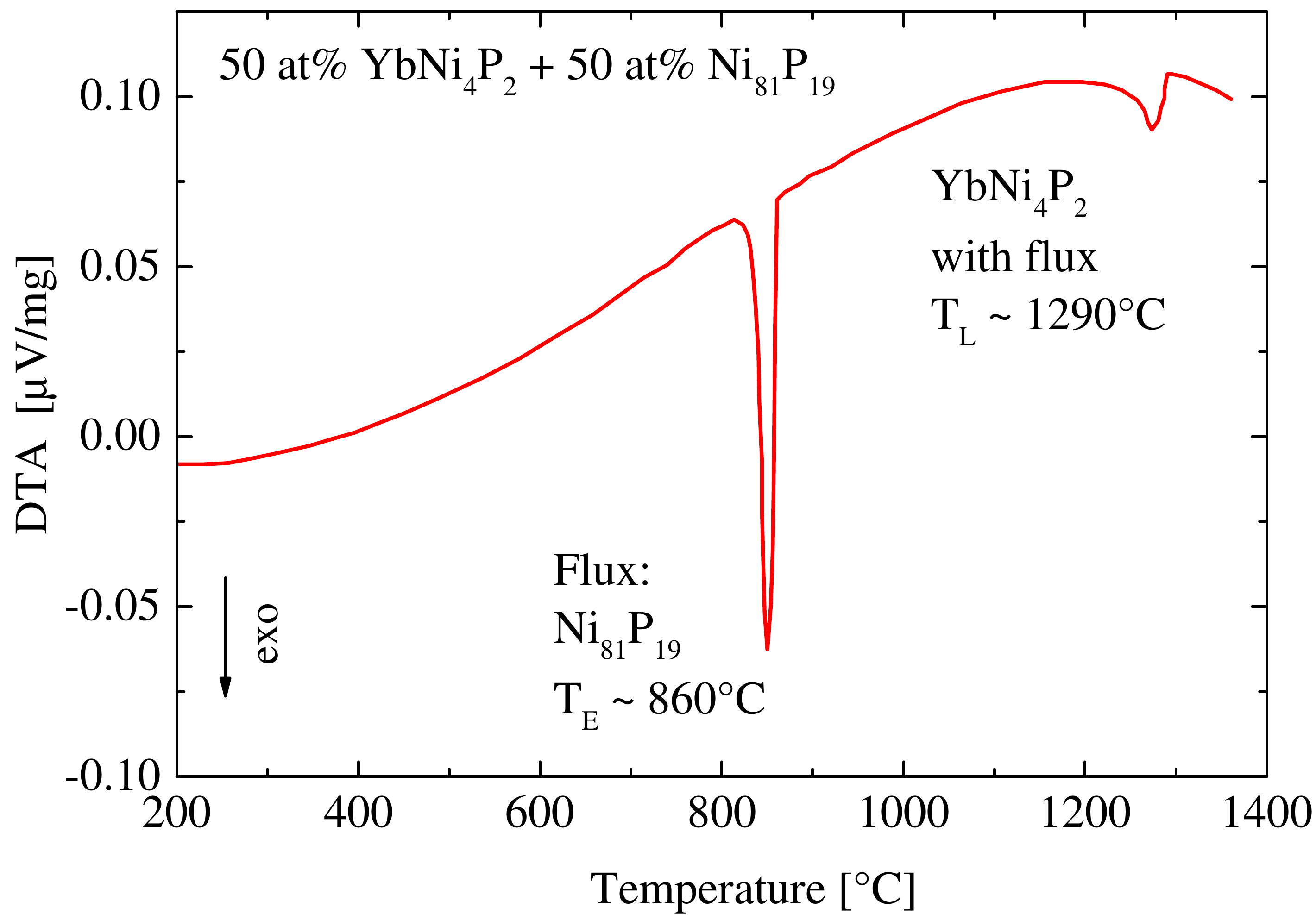}
\caption{The DTA signal recorded during cooling shows a dip at the liquidus temperature of the starting charge T$_{\rm L}\approx 1290^\circ$C marking the onset of the crystallisation of YbNi$_4$P$_2$. A second dip occurs at the eutectic temperature T$_{\rm E}\approx 860^{\circ}$C.}
\label{DTA}
\end{figure}
\begin{figure}[ht]
\centering
\includegraphics[width=0.32\columnwidth]{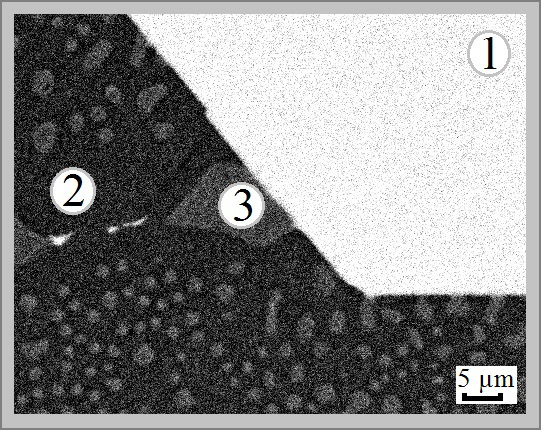}
\includegraphics[width=0.15\columnwidth]{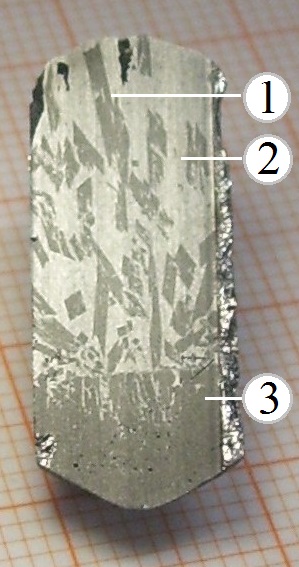}
\caption{Bridgman growth: {\it Left:} The secondary electron image of the cut through the sample shows YbNi$_4$P$_2$ crystals (1) in a matrix of Ni$_3$P (2) and Ni (3). {\it Right:} The YbNi$_4$P$_2$ single crystals (1) are enclosed in the Ni-P flux (2). A polycrystalline part of YbNi$_4$P$_2$ (3) formed at the bottom of the crucible. This axial cut through the ingot shows the distribution of phases that is expected from a directional solidification experiment according to the Bridgman technique. The ratio of the polycrystalline part and the part where the single crystals are surrounded by flux varied between different growth experiments.}
\label{eutectic}\label{cutimage}
\end{figure}
\begin{figure}[ht]
\centering
\includegraphics[width=0.5\columnwidth]{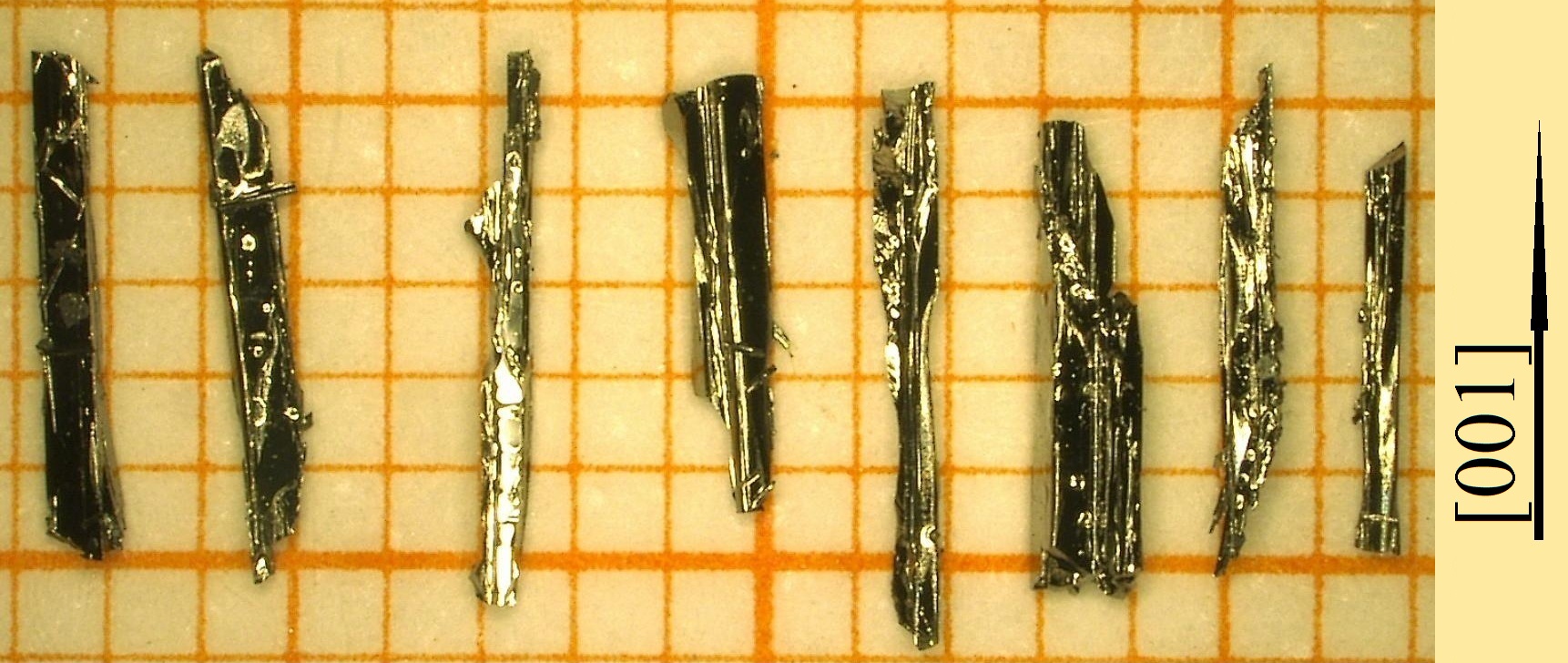}
\caption{Bridgman growth: Single crystals after the centrifugation procedure. The crystals grow preferentially along the $[001]$-direction.}
\label{Bridgmancrystals}
\end{figure}
\begin{figure}
\centering
\includegraphics[width=0.5\columnwidth]{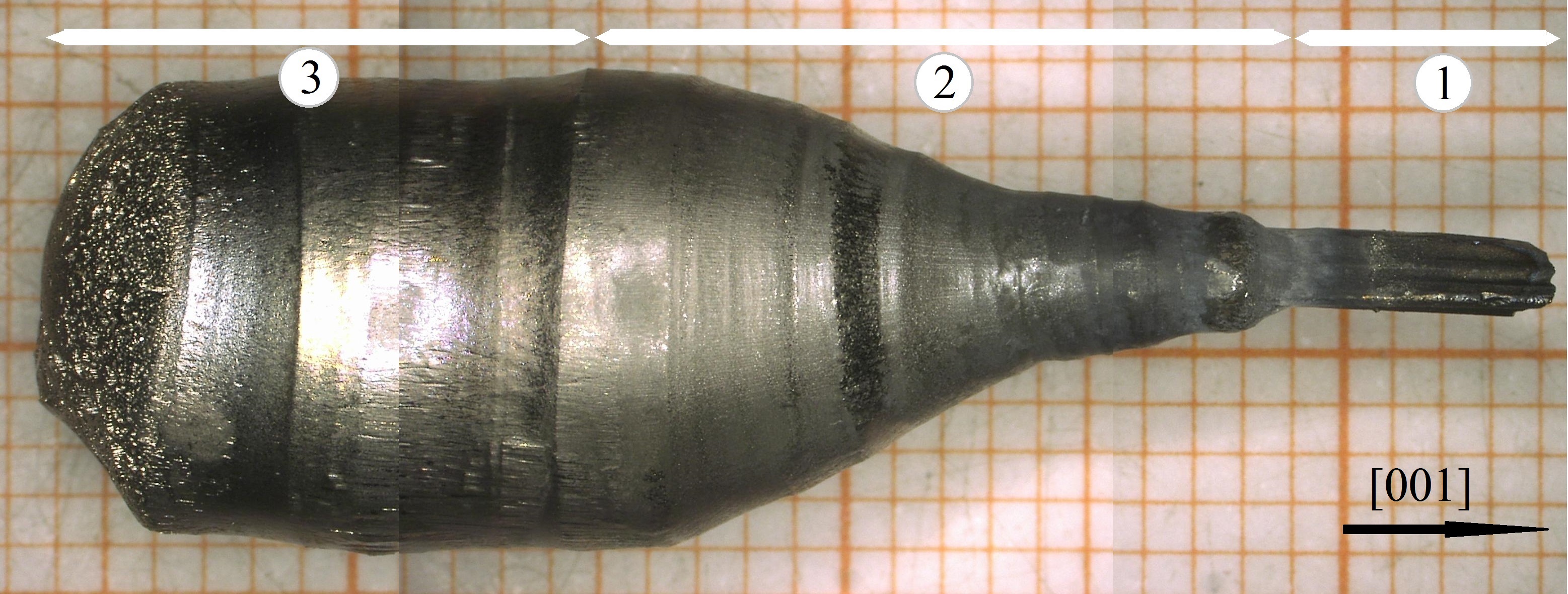}
\caption[]{Single crystal grown by the Czochralski method (Optical microscope image). An oriented seed (1) was used to grow an YbNi$_2$P$_2$ single crystal (2). The growth was terminated with a faster growth velocity which leads to an enhanced occurence of flux inclusions in the lower part of the sample (3). }
\label{sample039}
\end{figure}
\begin{figure}[ht]
\centering
\includegraphics[width=0.22\columnwidth]{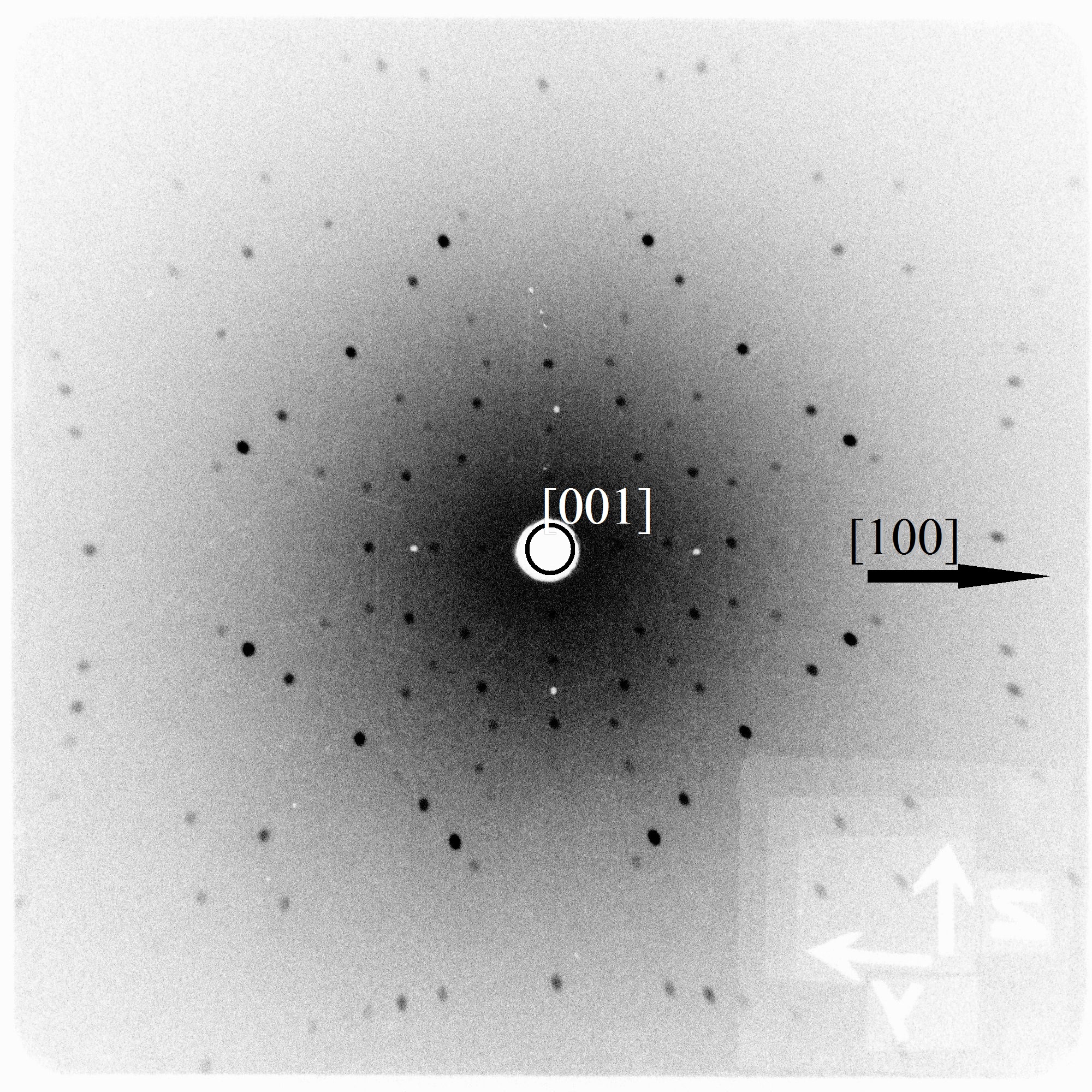}
\includegraphics[width=0.22\columnwidth]{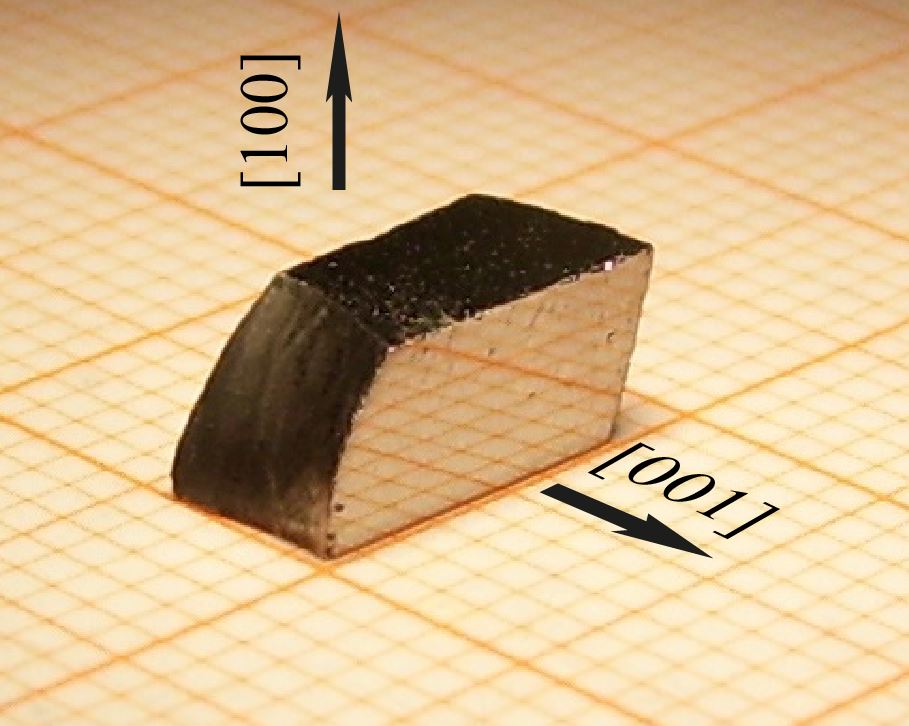}
\caption{{\it Left:} The Laue pattern of the [001]-direction of a sample prepared by Czochralski growth shows the four-fold symmetry. The good crystal quality is indicated by the sharp Laue reflexes. {\it Right:} Single crystal sample cut for a magnetization measurement.}
\label{laueCz}
\label{CzSample}
\end{figure}
\begin{figure}
\centering
\includegraphics[width=0.5\columnwidth]{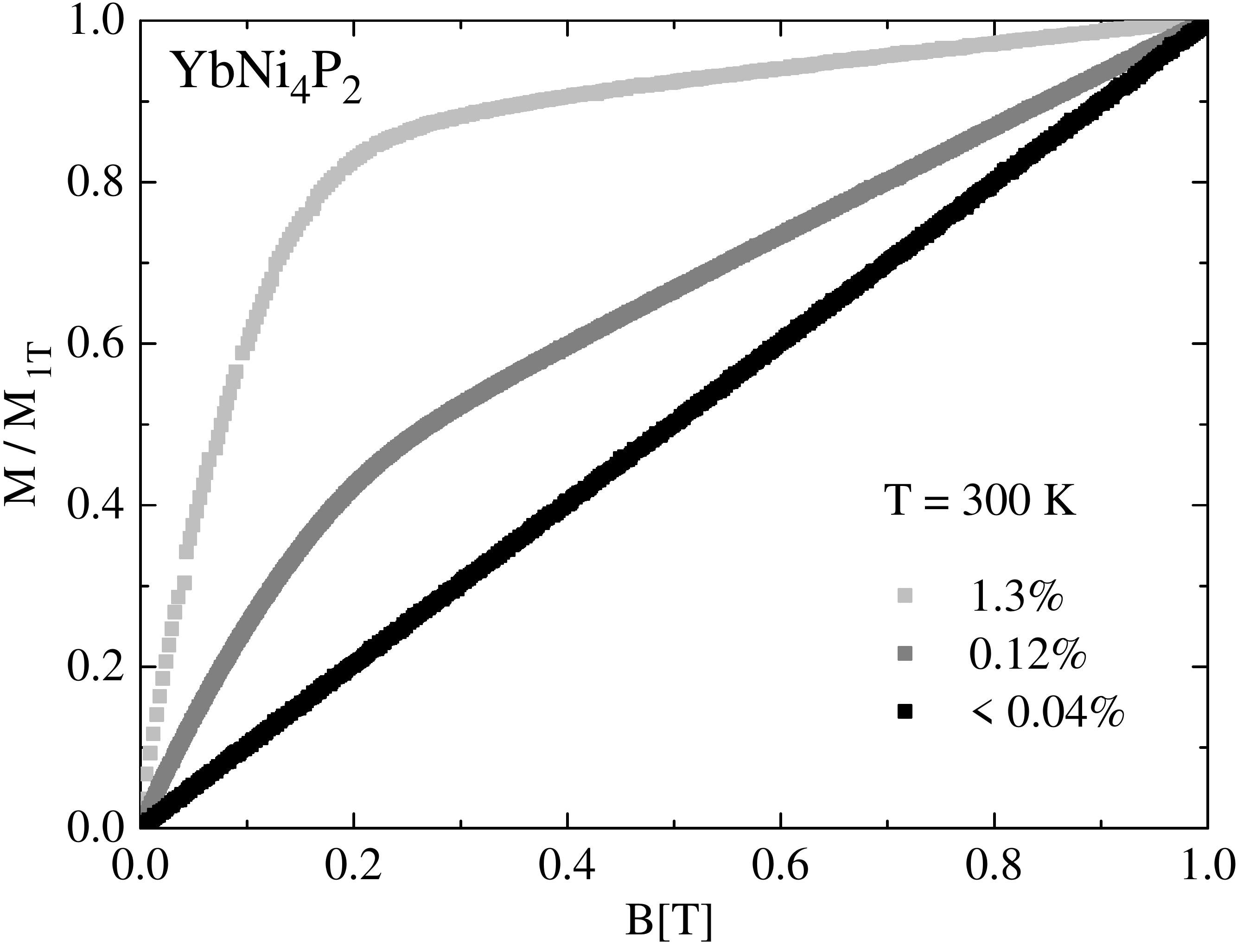}
\caption{Measured magnetization M(B) normalized to M(B = 1T) measured at T = 300 K on crystals with different Ni inclusions. 
The magnetic moment of YbNi$_4$P$_2$ depends linearily on B and is small at low fields. 
Below $\approx$ 0.1 T the measured moment is dominated by the contribution from the Ni inclusions. 
The black curve shows M(B) measured on a crystal with a Ni-content which is below the detection limit of this method. }
\label{nickel}
\end{figure}
\begin{figure}
\centering
\includegraphics[width=0.5\columnwidth]{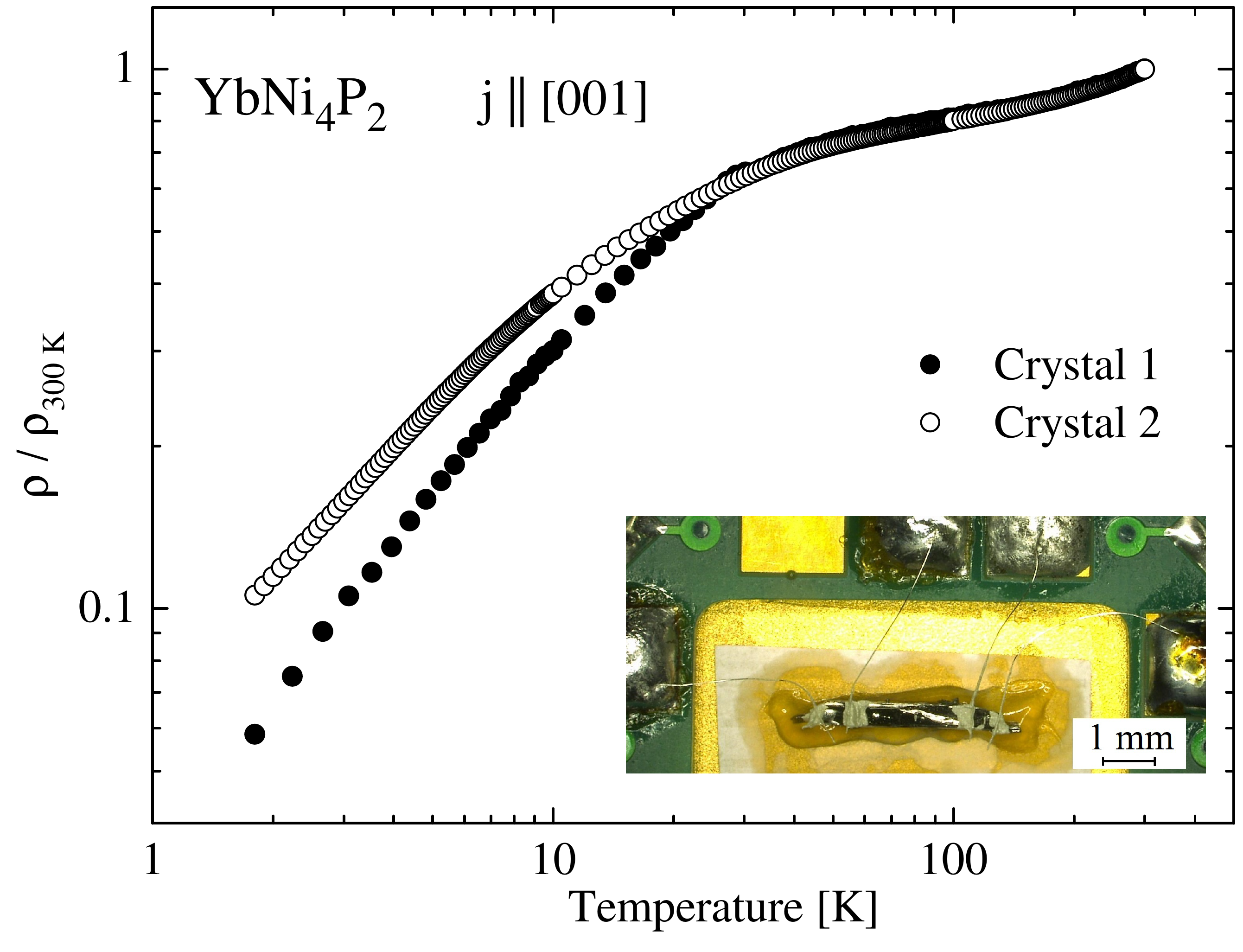}
\caption{Measured resistivity $\rho$(T) normalized to the resistivity at T = 300 K. RR$_{1.8 \rm K} = 17$ was determined for crystal 1  which was grown by the Czochralski method (closed symbols). 
For a crystal grown by the Bridgman method, crystal 2, we determined RR$_{1.8 \rm K} = 9$ (open symbols). 
In the inset, a needle shaped crystal connected to the sample platform in four-point-geometry with platinum wire contacts prepared for a resistivity measurement is shown.}
\label{RHO}
\end{figure}

\section{Experimental details}
High-purity starting materials Yb ingot (99.9\%, Strem Chemicals),
Ni slugs (99.995\%, Alfa Aesar), 
red P pieces (99.999\%, Mining \& Chemical Products Ltd.) were used. 
Some of the reagents, namely ytterbium and phosphorous, are air sensitive.
The preparation of these reagents was done in a glove box filled with purified argon.
The stoichiometric composition of the elements was weighed in together with 
50 at\% Ni$_{81}$P$_{19}$ (eutectic composition) as flux resulting in a sample to flux ratio of 1:1.
The total mass of each growth charge was 15 g.
The elements were filled in a graphite crucible (V = 25 ml)  
for the Bridgman growth and in a 
boron nitride crucible (V = 30 ml)  
for the preparation of the precursor for the Czochralski growth. The inner crucible
was put in an outer crucible made of tantalum which was sealed under Ar using arc-melting.
Differential thermal analysis was done using  
a Simultaneous Thermal Analysis device (STA 449 C, Netzsch), which allows simultaneous thermogravimetry (TG) 
and differential thermal analysis (DTA).
For the Bridgman growth, the Ta-crucible was put under a stream of Ar in a vertical resistive furnace 
(GERO HTRV70–250/18) in which a maximum temperature of 1350$^{\circ}$C was used in our experiment. 
During the growth, the temperature was measured in situ at
the bottom of the tantalum crucible by a thermocouple of type B.
After the Bridgman growth, the excess flux was spun of in a centrifuge (Christ UJ1) at $\approx 1100^{\circ}$C.
The Czochralski growth experiment was performed in a commercial ADL (Arthur D. Little) high-frequency growth device equipped 
with a generator that provides a maximum power of about $30~\rm kW$. The temperature was measured with an IRCON pyrometer.
The crystal structure was characterized by powder X-ray diffraction on
crushed single crystals, using Cu-K$_{\alpha}$ radiation.The chemical
composition was measured by energy-dispersive X-ray spectroscopy
(EDX). The orientation of the single crystals was determined using
a Laue camera with X-ray radiation from a tungsten anode. 
Four-point resistivity and magnetization measurements
were performed using the commercial measurement options of a
Quantum Design PPMS.


\section{Crystal growth}
A complete ternary phase diagram of Yb-Ni-P compounds at high temperatures
does not exist, but an isothermal section (T = 870 K) of this phase diagram
was determined by Kuz'ma et al. \cite{Kuzma2000}. Several stable ternary phases
exist in the vicinity of YbNi$_4$P$_2$. 
In previous work, the decomposition of YbNi$_4$P$_2$ above $1500^{\circ}$C
at ambient pressure was observed \cite{Krellner2011}. Therefore, one expects that the
crystal growth of the stoichiometric compound by floating-zone or the Czochralski method not to be successful. 
The binary Ni-P phase diagram shows a low-melting eutectic, Ni$_{81}$P$_{19}$ \cite{Okamoto2010}. A detailed investigation identified Ni$_{80.4}$P$_{19.6}$ as the eutectic composition with the eutectic temperature T$_{\rm E}=875^{\circ}$C \cite{Huang2010}. 
We have used this eutectic as a self-flux in one series of experiments utilizing a Bridgman 
and in an other series the Czochralski technique to grow
YbNi$_4$P$_2$ single crystals. 
One further problem is that the Yb-Ni-P melt exhibits a high reactivity with other materials leading to lack of inert crucible material.  
For the determination of the crystallisation temperature of YbNi$_4$P$_2$ in Ni$_{81}$P$_{19}$ simultaneous
TG and DTA have been performed before starting the growth experiments.
555 mg of prereacted material consisting of 50~at\% YbNi$_4$P$_2$ 
and 50~at\% Ni$_{81}$P$_{19}$ was put in an open alumina crucible and heated with 10 K/min in an Ar stream.
The weight loss after 3 heat/cool cycles was $\Delta m/m\approx 0.05$ and the signals of all three runs were reproducible. 
The DTA curve presented in Fig.~\ref{DTA} shows the third cooling process. 
During heating, the melting signal of the eutectic shows up at 870$^\circ$C. 
The melting signal of the 142-compound is located at $\approx$1340$^\circ$C and relatively weak.
The high melting temperature of the transition metal Ni (1455$^{\circ}$C) in combination 
with the starting sublimation of red P at low temperature (416$^{\circ}$C)
and its high reactivity additionally to the low boiling point (1196$^{\circ}$C) 
and high vapour pressure of Yb necessitates the preparation of a precursor.

\subsection{Bridgman method}

YbNi$_4$P$_2$ single crystals were grown by a modified Bridgman method
from a Ni-P self-flux for the first time in 2012 \cite{Krellner2012}. 
In the mean time, the growth procedere has been optimized and several physical properties of this compound have been investigated, but a detailed description of the growth parameters has not been reported yet.
For the Bridgman growth, 
the sealed Ta-crucible was slowly heated up to 700$^{\circ}$C 
with a rate of 30 K/h to allow a slow reaction of phosphorous with the other elements 
and to 1350$^{\circ}$C with a rate of 50 K/h.
The melt was held at this temperature for 1 h to ensure homogenization and then cooled by slow moving of the whole
furnace with 0.88 to 3.4 mm/h leading to a cooling rate in the range of
0.5 - 4 K/h down to 1000$^{\circ}$C, while the position of the crucible
stayed fixed. With this setup, we are able to cool the sample without vibrations resulting from the movement 
which is different from the conventional Bridgman process where the sample is moved from the hotter to the cooler zone.
After the growth, the YbNi$_4$P$_2$ single crystals are embedded in the Ni-Ni$_3$P eutectic. 
A typical growth result (cut image) with 
the YbNi$_4$P$_2$ single crystals embedded in the flux
 is shown in  Fig.~\ref{cutimage}.
Since the flux can not be removed by acids without solving the crystals, the use of a centrifuge was necessary to separate the crystals from the flux.
In preparation of the centrifugation process, the sample was cut using a spark erosion device and placed above some glassy carbon pieces 
and a graphite sieve in a fused silica ampoule.
The ampoule was heated in a box furnace up to 1100$^\circ$C, held at this temperature for one hour 
and then within a few seconds moved into a centrifuge. The flux with the eutectic temperature of 
$\approx$ 870$^\circ$C was spun of.
Afterwards, the remaining crystals could be easily separated from each other manually. The long, rod shaped single crystals are presented in Fig.~\ref{Bridgmancrystals}.


\subsection{Czochralski growth from a levitating melt}
In the past, the successful single crystal growth of cerium compounds in the same high frequency furnace that we used  
has been reported 
\cite{Nuettgens1997,Takke1980,Nuettgens2000}. 
YbNi$_4$P$_2$ single crystals were grown from a levitating melt applying the Czochralski method using the same sample to flux
ratio of 1:1 that was used in the Bridgman experiments described above.
A precursor was prepared using a boron nitride crucible, 
 welded inside a tantalum crucible using an argon arc furnace  
and were prereacted in a box furnace under argon
atmosphere subsequently. 
The Czochralski growth experiment was started by melting the precursor material in a cold copper crucible (Hukin-type) 
with a radio-frequency induction coil applying a power of $12~\rm kW$ at maximum.  
The precursor was homogenized due to the strong stirring of the levitating melt within several minutes. 
The power was set in a way that the temperature of the melt was above the liquidus temperature 
at about 1400$^{\circ}$C. The melt was kept at this temperature for 15 min to ensure complete homogenization.
For the first Czochralski growth, we used an YbNi$_4$P$_2$ seed prepared from a Bridgman-grown crystal. 
In the following experiments, oriented seeds prepared from the first Czochalski experiment were used.
The seed was lowered into the melt and the generator power was adjusted carefully after dipping. 
As soon as the process run stable after dipping, the seed was pulled upwards along its crystallographic $[001]$-direction
with a pulling rate of $0.2~\rm mm/h$. Within the process time of $30~\rm h$ the total power reduction was about 30\% during the experiment. 
To achieve an inclusion free sample, a low growth rate and 
a long process time of several days was necessary.
First experiments applying an argon pressure of $2~\rm bar$ led to considerable evaporation of phosphorous from the melt 
and therefore to a shift of the stoichiometry which made the crystal growth unstable.
This evaporation of phosphorous from the melt 
was slowed down by applying an argon pressure of $20~\rm bar$ in the growth chamber leading to stable growth conditions.
Fig.~\ref{sample039} shows a typical growth result.
\section{Sample characterization}
\subsection{Structural and chemical characterization}
Powder X-ray diffraction measurements confirmed the 
tetragonal ZrFe$_4$Si$_2$ structure type (P$4_2$/mnm) with lattice parameters  
a=7.0560(3)\AA\,  and c=3.5876(5)\AA , which are in agreement with the data published 
for polycrystalline samples \cite{Kuzma2000}.
EDX microprobe analysis of the crystals grown by the two different methods, revealed the stoichiometry of the 142-compound within an error of 2~at\%. 
The single crystals additionally were analyzed with electron microscopy.
The cut through a Bridgman growth sample showed that 
YbNi$_4$P$_2$ single crystals were embedded in the Ni-Ni$_3$P eutectic 
(Fig.~\ref{eutectic}). 
The high quality of the single crystals is evident from sharp Laue back scattering spots shown in Fig.~\ref{laueCz}. 
These samples sometimes exhibit inclusions of the eutectic. Besides the para-magnetic \cite{Gambino1967} Ni$_3$P phase, the eutectic flux also contains small inclusions of magnetic Ni.

\subsection{Magnetization and electrical transport measurements}
The grown YbNi$_4$P$_2$ crystals can contain flux inclusions consisting of the
eutectic mixture of Ni$_3$P and Ni with the phase fractions of approximately 3:1 
according to the lever rule. 
While Ni causes a ferromagnetic contribution to the magnetization of the sample,
Ni$_3$P is paramagnetic.
To estimate the residual Ni content, magnetic measurements were performed, Fig.~\ref{nickel}. With this very sensitive method, 
the purest crystals concerning the flux content can be identified since 
the contribution of nickel to the measured moment is large and reaches 90\% of the saturation at low fields 
B $\approx$ 0.1 T. We utilized the fact that the contribution of the Yb moments increases linearily, and 
is therefore small at lower fields. 
As shown in Fig.~\ref{RHO}, electrical transport measurements on YbNi$_4$P$_2$ single crystals with current
along the $[001]$-direction were performed 
to compare the crystal quality of samples 
from different batches by means of the resistivity ratio $RR_{1.8\rm K}:= \rho(300\rm K)/\rho(1.8\rm K)$.

\section{Results and discussion}

\subsection{Crystal growth by Bridgman method}

The YbNi$_4$P$_2$ single crystals grown by the Bridgman method exhibit naturally grown faces. The rod-shaped crystals grow preferentially along the $[001]$-direction, 
whereas the tetragonal plane is bounded by $\{110\}$ faces.
The dimensions of the largest crystals were $\rm 0.8 mm \times 0.9 mm \times 6 mm$.
During the growth, the melt attacked all tested crucible materials (Al$_2$0$_3$, glassy carbon, graphite, tantalum) leading to a contamination of the melt. 
The analysis by carrier gas hot extraction of a polycrystalline sample showed that the crystals grown in a graphite crucible contain up to 1 wt\% carbon. 
The crystals grown by the Bridgman method sometimes contain inclusions of residual flux consisting of the Ni-Ni$_3$P eutectic. 
This content of flux was estimated by magnetic measurements to be 0 - 1 wt\%. 
The largest Ni-free single crystals have a mass of about 10 mg.
We performed electrical-transport measurements with current parallel to the crystallographic $[001]$-direction and 
determined  RR$_{1.8 \rm K}=9$ for the best crystals from the Bridgman growth experiments for current along the $[001]$-direction.

\subsection{Czochralski growth from a levitating melt}
Fig.~\ref{sample039} shows an example of a successful Czochralski growth from a levitating melt with a diameter of 9-10 mm and a total length of 14-15 mm.
The upper part of the sample (region 2) is about 9 mm long and consists of YbNi$_4$P$_2$. The mass of this single crystal part is about 1.8 g.
In the lower part of the sample, YbNi$_4$P$_2$ contains an increasing amount of flux inclusions (region 3). 
The single crystal seeds that were used have been oriented and all samples were pulled out of the melt along the $[001]$-direction.
A Laue image of a Czochralski grown single crystal is shown in Fig.~\ref{laueCz}.
Flux inclusion consisting of the Ni-Ni$_3$P eutectic have been observed in the lower part of the grown sample. These also occur in the upper part at grain boundaries in samples that contain more than one grain.
This content of flux was estimated by magnetic measurements to be 0 - 0.04 wt\%. Fig.~\ref{CzSample} shows a typical sample prepared for a magnetization measurement. 
We performed electrical-transport measurements with current parallel to the crystallographic $[001]$-direction
and determined  RR$_{1.8 \rm K}=17$ for the best crystals from the Czochralski growth experiments.
We found that with the crucible-free Czochralski method it is possible to grow large, inclusion-free single crystals with a higher residual resistivity ratio. 
The good quality of the samples became apparent since it was possible to observe quantum oscillations in the torque measured on the samples in fields above 20 T (to be published elsewhere \cite{Friedemann2016}).
\section{Summary}
YbNi$_4$P$_2$ single crystals have been grown by two different methods. 
The Bridgman method yields rod shaped up to 6 mm long single crystals. The mass of one crystal is 10~mg at maximum.
The preferred growth direction is the $[001]$-direction. The crystals exhibit naturally grown $\{110\}$ faces. 
The Czochralski method yields single crystals with masses up to 1.8~g which are up to 15 mm in length. With this crucible-free method inclusion-free samples with a higher residual resistivity ratio RR$_{1.8\rm K}= 17$ have been grown. 
In future work, it might be possible to further improve the crystal quality by annealing the crystals after the growth.
With this work we have shown that the high-temperature metal-flux technique is a powerful tool to grow large single crystals of materials with volatile elements. 
\section{Acknowledgements}
We thank C. Geibel, P. Gille, J. Schwerin, W. Assmus and F. Ritter 
for valuable discussions, G. Auffermann for the sample analysis by carrier gas hot extraction 
and K.-D. Luther for technical support.  


\section{References}



\bibliographystyle{unsrt}
\bibliography{YbNi4P2_Kliemt_final.bbl}



\end{document}